\documentstyle[twocolumn,psfig,aps]{revtex}
\begin{document}
\draft
\twocolumn[\hsize\textwidth\columnwidth\hsize\csname @twocolumnfalse\endcsname

\title{ Doped Stripes in Models for the Cuprates \\
Emerging from the One-hole Properties of the Insulator }

\author{G. B. Martins$^1$, C. Gazza$^2$, J. C. Xavier$^1$, 
A. Feiguin$^1$, and E. Dagotto$^1$}

\address{$^1$ National High Magnetic Field Lab and Department of Physics,
Florida State University, Tallahassee, FL 32306, USA}

\address{$^2$ Instituto de F\'{\i}sica Rosario (CONICET) and Univ. Nac. 
de Rosario, Bv. 27 de Febrero 210 bis, 2000 Rosario, Argentina}

\date{\today}
\maketitle

\begin{abstract}
The extended and standard t-J models are 
computationally studied on ladders and planes, with emphasis
on the small J/t region.
At couplings compatible with photoemission results for undoped
cuprates, half-doped stripes  
separating $\pi$-shifted antiferromagnetic (AF) domains are found,
as in Tranquada's interpretation of
neutron experiments. 
Our main result is that the 
elementary stripe ``building-block'' resembles the 
properties of $one$ hole
at small J/t, with robust AF correlations across-the-hole
induced by the local tendency of the charge to separate from the spin
(G. Martins {\it et al.},
Phys. Rev. B{\bf 60}, R3716 (1999)).
This suggests that the seed of half-doped stripes
already exists in the unusual properties of the
insulating parent compound.
\end{abstract}
\vskip2pc]
\narrowtext

\medskip

The understanding of high temperature superconductors 
is among the most important open problems in
strongly correlated electrons.
A remarkable development in recent years is the
accumulation of experimental evidence compatible
with stripe formation in the normal state of
underdoped cuprates\cite{tranquada}. This includes spin
incommensurability (IC) in neutron experiments,
results believed to be caused by stripes
separating $\pi$-shifted AF domains\cite{tranquada}.
More recently, it has been shown that the stripes are
metallic\cite{uchida}, result compatible with proposals
of the normal state of x=1/8 cuprates as made out
of half-doped stripes\cite{tranquada}.
Whether stripe formation is beneficial or detrimental to
superconductivity is unclear, but it appears
that stripes are an important ingredient of the
normal state that cannot be ignored.

The theoretical explanation of stripe formation is 
much debated. Early work reported stripes 
in the t-J (at large J/t with 1/r repulsions) 
and Hubbard
(Hartree-Fock) models\cite{zaanen,emery}. However, these stripes were
insulating with
hole density n$_h$$\sim$1.0, different from
the experimental n$_h$$\sim$0.5 stripes\cite{tohyama}.
Recently, considerable progress was made when 
doped stripes were reported by White and Scalapino
within the standard t-J model\cite{white} 
(see also Ref.\cite{moreo}).  
In Ref.\cite{white} the analysis
was performed at couplings where two holes 
form d-wave pairs, and the stripes are sometimes described as a
condensation of these pairs into a stripe domain-wall\cite{white2}.
However, experiments are usually interpreted as holes moving
freely along site-centered stripes\cite{tranquada}.
In addition, the ``extended'' t-J model 
with hopping beyond
neighboring sites, or the standard t-J model with very
small J/t, are needed\cite{martins,eder} to reproduce 
the insulator one-hole photoemission (PES) dispersion\cite{arpes}. 
Thus, understanding metallic stripe formation requires further work
and searching for stripes in the extended t-J model, particularly in
regimes without hole binding and where the absence of phase separation (PS) is not
controversial, is important
to clarify the driving mechanism for these unusual complex structures.

Building upon previous investigations\cite{white,moreo},
in this Letter indications of n$_h$$\sim$0.5 stripes 
compatible with experiments\cite{tranquada} are reported
in the extended and standard t-J models on
ladders and square clusters. These stripes do not seem composed
of hole pairs (although pairs forming domain-walls may be present
at larger J/t than studied here\cite{white2}). 
They also exist in the t-J$_z$ model\cite{castroneto}
and using classical spins\cite{moreo}, implying
that the details of the AF spin background are unimportant for its
stabilization. Moreover, our most important result 
is that the basic stripe ``building-block'' 
exists already in the {\it insulator} where one-hole wave functions 
have a complex spin structure with strong AF correlations
across-the-hole, resembling 
the stripe spin correlations found here numerically. These results 
provide a rationalization for stripe formation 
built upon the $one$ hole properties, in regimes where
spin and charge are almost separated\cite{martins}.

The extended t-J model used here is defined as
$$
\rm
H = J \sum_{\langle {\bf ij} \rangle} 
({{{\bf S}_{\bf i}}\cdot{{\bf S}_{\bf j}}}-{{1}\over{4}}n_{\bf i}n_{\bf j})
- \sum_{ {\bf im} } t_{\bf im} (c^\dagger_{\bf i} c_{\bf m} + h.c.),
$$
where $\rm t_{\bf im}$ is t(=1) for nearest-neighbors (NN), t$'$ for 
next NN, and t$''$ for next next NN sites, and
zero otherwise. The rest of the notation is standard. The t-J$_z$ model
is obtained by J$\rightarrow$J$_z$ and
$\rm {{{\bf S}_{\bf i}}\cdot{{\bf S}_{\bf j}}}$$\rightarrow$
$\rm {{S^z}_{\bf i}}{{S^z}_{\bf j}}$,
and t$'$$<$0 and
t$''$$>$0 are relevant to explain PES data\cite{martins,eder,arpes}.
Here the Density Matrix Renormalization Group (DMRG)\cite{white,error},
Lanczos\cite{review}, and 
an algorithm using
a small fraction of the ladder rung-basis 
(optimized reduced-basis approximation, or ORBA\cite{orba}) are used. 
Results are presented in (i) the
small J/t region with t$'$=t$''$=0.0, and (ii) small
and intermediate J/t with nonzero t$'$ and t$''$\cite{eder}. These two
regions have similar physics\cite{martins}, and the extra
hoppings are expected to avoid
PS\cite{tohyama,contro}.
Intuitively, t$'$,t$''$
increase hole mobility, as reducing J/t does,
but also avoid ferromagnetism at small J/t\cite{martins}.
Note also that no coupling fine-tuning is needed: the results below
appear in a robust region of parameter space. 

\vspace{-0.2in}

\begin{figure}
\begin {center}
\mbox{\psfig{figure=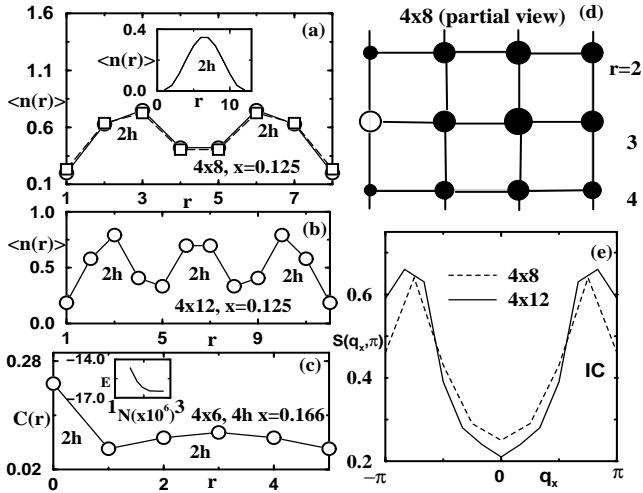,height=2.5in,width=3.2in,rheight=2.5in,rwidth=3.5in,angle=-90}}
\end{center}
\caption{(a,b) Rung hole density $\langle$n(r)$\rangle$ vs rung index r
using DMRG, with PBC along rungs 
and OBC along legs, to
illustrate the n$_h$$\sim$0.5 stripe formation. 
x is the overall hole density.
(a) corresponds to a 4$\times$8 cluster with 4 holes. 
Solid (dashed) lines are for the standard t-J (t-J$_z$) model with
J=0.2 (J$_z$=0.3), 
t$'$=t$''$=0.0 (inset: same as solid lines but for 4$\times$12 with 
2 holes). (b) Same as (a) but for a 
4$\times$12 cluster with 6 holes, J=0.5, t$'$=-0.3, and 
t$''$=0.0. (c) Hole density at rung r, defined now as
$\rm C(r)$=$\rm \sum_{i \epsilon r} \langle n_{\bf 0} n_{\bf i} \rangle$
where the sum is over sites belonging to  rung r, ${\bf 0}$ is an 
arbitrary site of rung r=0, and $\rm \langle n_{\bf 0} n_{\bf i} \rangle$
is the hole density-density ground-state correlation. 
The cluster is 4$\times$6 with PBC in both
directions, 4 holes, J=0.2, t$'$=-0.35, and t$''$=0.25 (ORBA
with $\sim$3$\times$10$^{6}$ states).
The inset shows ground-state energy vs
number of states. 
(d) Distribution of one-hole around a second hole
projected at the open circle position, for the case in (a) at
the indicated rungs (running horizontally). Full circles areas are
proportional to the hole density. 
(e) S(q$_x$,$\pi$) vs 
q$_x$ for the clusters, couplings, and densities
of (a) and (b).
} 
\label{fig1}
\end{figure}

\vspace{-0.2cm}

In Fig.1, DMRG and ORBA results for 4$\times$N clusters are shown.
In Fig.1a the rung density
for a 4$\times$8 (4$\times$12) cluster with 4 (2) holes at
small J/t is presented. Cylindrical
 boundary conditions (CBC) are used i.e. open boundary conditions (OBC) 
along legs and periodic boundary conditions (PBC) 
along rungs\cite{white}.
The four holes separate into two groups of two holes, 
surprising result since for
a square lattice J$_c$=0.2 is the critical value for
hole pair binding in the t-J$_z$ model, and in the t-J model J$_c$ is
expected to be larger\cite{jose}.
Similar results
are found in the t-J$_z$ model (Fig.1a)
and at intermediate J/t but with t$'$$\neq$0,
which increases the hole mobility: 
Fig.1b with six holes show the formation
of three groups of two holes as in Fig.1a. This is not 
spuriously caused by the
OBC along legs, as shown in Fig.1c with results
using PBC in both directions. 
As ORBA starting
configuration holes clustered (phase separated) or 
spread apart (free gas)
were used, with PBC or CBC, 
and in both cases the results converged to the same ``stripe'' 
answer. 

To study the two-hole state internal structure, 
in Fig.1d the density distribution of one hole around 
the other is shown, for one of the 2-hole regions of Fig.1a.
The largest density is at two lattice spacings
along the rung, 
and the hole distribution does not resemble a tight d-wave bound
state\cite{review}.
Similar conclusions were reached for the two holes 
of Fig.1c. 
The result is actually compatible with the formation of a short 
site-centered stripe where the two holes 
form a closed loop with density 0.5
along a rung\cite{comm10}.
These stripes appear to occupy more than one rung in Figs.1a-c,
and thus they could also be labeled as bond-centered\cite{white}. However, 
this effect seems to arise from stripe 
tunneling between neighboring rungs, as the
one-hole projection suggests (Fig.1d).
Similar results regarding half-doped 
stripe formation were also found on 6$\times$6 
clusters, as exemplified in Fig.2a where sets of 3 holes form individual
n$_h$$\sim$0.5 stripes (invariance under reflexions 
was assumed along the legs). Overall the results are
consistent with Tranquada's description of stripes\cite{tranquada}. 
They are also consistent with 
numerical reports for the standard t-J model\cite{white}, although
our interpretation of the results (below) is different.

\begin{figure}
\begin {center}
\mbox{\psfig{figure=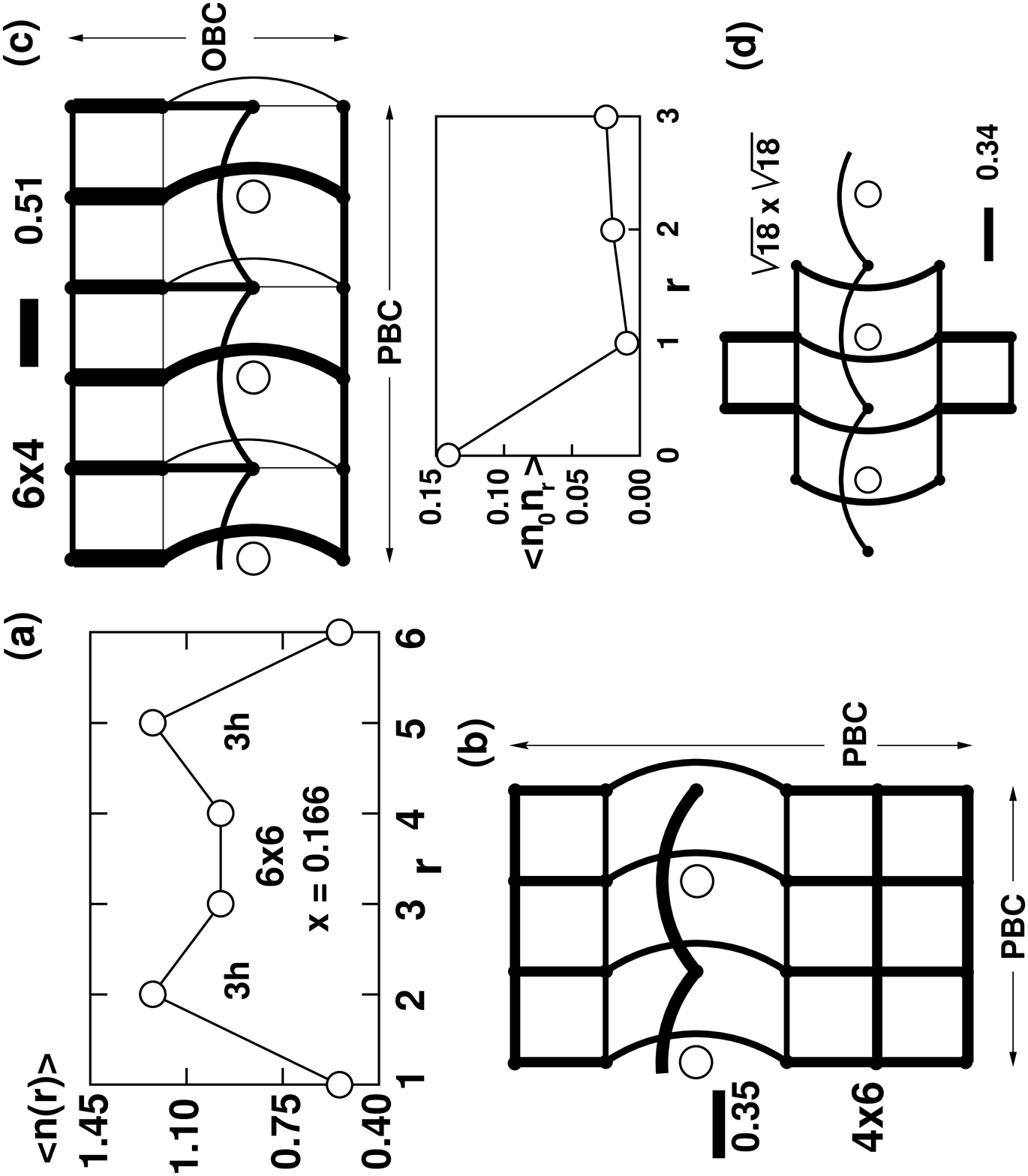,height=2.5in,width=3.in,rheight=2.5in,rwidth=3.2in,angle=-90}}
\end{center}
\caption{(a) Rung hole density $\langle$n(r)$\rangle$ vs rung index r
using DMRG on a 6$\times$6 cluster with 6 holes, 
CBC (OBC along the direction shown with invariance under reflexions
assumed),
J=0.2, and t$'$=t$''$=0.0 (600 states, 8 sweeps).
(b) Spin-spin correlations for 2 mobile holes projected at their
most probable
relative distance (circles) 
in the 2 holes ORBA ground state
of a 4$\times$6 cluster, 
J=0.2, t$'$=-0.35, and t$''$=0.25. Lines
indicate AF correlations (thickness proportional to
absolute value).
(c) Same couplings, cluster, technique, and conventions
as in (b) but using PBC (OBC) along legs (rungs), and
3 holes. Also shown are the hole density-density correlations along 
a center leg, showing that there is no charge order.
(d) AF spin correlations 
for a large weight ground-state configuration
of an exactly solved 3 hole, 18 site PBC cluster,
J=0.4, t$'$=-0.20 and t$''$=0.14.
} 
\label{fig2}
\end{figure}

\vspace{-0.2cm}

The half-doped stripes reported here also
lead to spin IC. For example, in Fig.1e
the spin structure factor
is shown for the cases of
Figs.1a,b. The peak deviation from 
($\pi$,$\pi$)
appears in a robust region of parameter space. 
The spin IC is understood calculating
spin-spin correlations when two holes in, e.g., the
cluster of Fig.1c are projected into their most probable location 
(Fig.2b): a 
$\pi$-shift across-the-stripe can be clearly observed.  
The across-the-stripe AF correlation strength increases
reducing J/t and/or increasing t$'$$<$0 and 
t$''$$>$0 in magnitude.

Results compatible with n$_h$$\sim$0.5 stripes and associated $\pi$-shifts  
appear in other clusters as well.
On a cylindrical
6$\times$4 cluster with PBC along the long direction, 
the 3-holes ground-state has
characteristics compatible with a doped one-dimensional (1D)
closed loop along the PBC direction,
with $\pi$-shifts across-the-stripe (see
Fig.2c where one of the two
degenerate most dominant ground-state
hole configurations is shown). 
A h-s-h-s-h-s loop (h=hole, s=spin) provides a pictorial
representation of our results,
but this configuration is not rigid neither
along nor perpendicular to the loop.
Density correlations along the stripe (Fig.2c) 
are actually compatible with a 1D n$_h$$\sim$0.5 system at large on-site
U interactions\cite{shiba}, suggesting that
the stripes described  here are metallic. No indications of a
charge-density-wave along the stripe were found.
Note also that spin IC induced
by antiferromagnetism across holes also exist $along$ $the$ $stripes$, 
with wavevector $\pi$/2 for a half-doped stripe. This spin IC
appears also in half-doped 1D models\cite{shiba}. For an isolated 
CuO plane, IC should be present in both
directions, although with quite different wavevectors and intensities.

Similar results are found in small square clusters: in
the 2-holes 4$\times$4 lattice with CBC,
a 2-hole stripe forms along the
PBC direction\cite{martins}.
With PBC in both directions, the ground-state resembles
a mixture of stripes along both axes and since
nonzero t$'$-t$''$ avoids PS, our
results are not expected to have the boundary effects recently 
discussed\cite{contro}.
Tendency to stripe formation is found
even in tilted clusters:
the PBC $\sqrt{18}$$\times$$\sqrt{18}$
lattice allows for n$_h$$\sim$0.5 closed loops with 3 holes 
and such structure has a large ground-state weight (Fig.2d)\cite{diagonal}.
Precursors of the spin structures in Figs.1,2 
appear on 2- and 3-leg ladders as well, e.g.
in Fig.3a the 2 holes ground-state 
dominant hole configuration of a 3$\times$6
cluster is shown, with its spin correlations.
On 2-leg ladders with many holes,
$\pi$-shifts appear at small J/t (Fig.3b), and each
hole is ``confined'' to a rung, precursor of a rung stripe
as the leg number grows. 
Spin IC is here found both for the 2-leg (Fig.3c) and 3-leg ladders.

The results thus far suggest that doped stripes
can form in spin and hole models using realistic couplings.
To gain insight into the mechanism driving this complex structure, 
consider now the $one$ hole problem. Fig.3d shows 4-leg ladder 
spin correlations around a mobile hole for momentum ($\pi$,$\pi$). 
The AF correlations across-the-hole are clearly 
similar to the correlations around
the individual holes composing the stripes. 
The $\pi$-shift characteristic of the stripes exists in the
one-hole state not only at ($\pi$,$\pi$) but at several momenta,
and, in this sense, the spin IC exists already
at the one-hole level, a remarkable result. 
Similar conclusions
are reached for 3- and 2-leg ladders (Fig.3e), and other 
momenta such as (0,$\pi$).  Also on small square clusters
robust across-the-hole AF correlations exist for one hole. 
Although 
spin IC was found in early studies of the t-J model\cite{review},
and the nontrivial structures as in Fig.3d were noticed before\cite{white},
it was only recently tentatively explained\cite{martins}
as (local) spin-charge separation, similar to the 1D 
Hubbard model where spins across holes are antiparallel\cite{shiba}. 

\vspace{-0.1in}

\begin{figure}
\begin {center}
\mbox{\psfig{figure=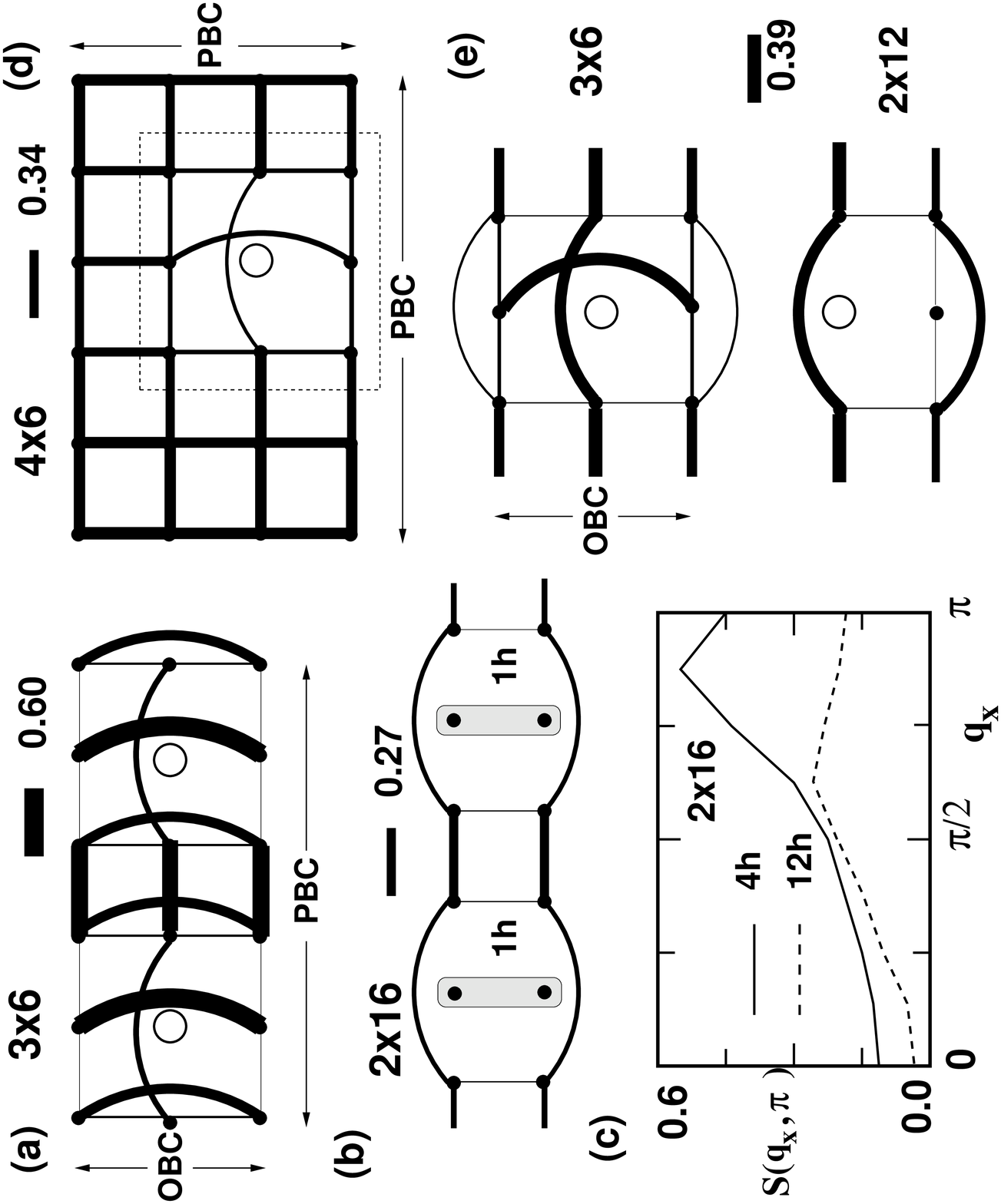,height=2.5in,width=3.in,rheight=2.5in,rwidth=3.2in,angle=-90}}
\end{center}
\vspace{-0.2cm}
\caption{
(a) AF spin correlations for 2 holes on
the 3$\times$6 cluster solved exactly 
at J=0.2, t$'$=-0.35, t$''$=0.25,
with holes projected at their most probable distance in the ground state.
(b) AF spin correlations at the center of a 2$\times$32 cluster with 12 holes
(OBC-legs) using DMRG at J=0.2, t$'$=t$''$=0.0.
Shaded regions contain the holes.
(c) S(q$_x$,$\pi$) vs q$_x$ 
for a DMRG 2$\times$16 cluster with 
12 and 4 holes and couplings (J=0.4, t$'$=-0.35, t$''$=0.25) and
(J=0.2, t$'$=t$''$=0.0), respectively.
(d) Exact AF spin correlations 
of the PBC 4$\times$6 cluster with one hole and
$\rm {\bf q}$=($\pi$,$\pi$), J=0.2, t$'$=-0.35, and t$''$=0.25.
An elementary block conjectured to form part of
stripes is framed.
(e) Results for 1 hole as in (d) but for 
a 2$\times$12 cluster at J=0.2, and a 3$\times$6 cluster at J=0.1, both
for t$'$=-0.35, t$''$=0.25 and $\rm {\bf q}$=(0,$\pi$).
} 
\label{fig3}
\end{figure}

\vspace{-0.4cm}

The results shown here lead us to believe that the
observed doped stripes are made out of 
one-hole  building-blocks (Fig.3d). 
In this respect the insulator limit 
already carries the essential information
needed to build the stripes, providing an unexpected potential
simple link between undoped and doped cuprates. This is compatible with
the behavior of the large energy scale PES pseudogap
which can be traced back to the
one-hole dispersion of the insulator\cite{arpes}, suggesting a smooth
evolution from the undoped to underdoped regimes.

However, further elaboration is needed
since for one-hole the lowest
energy is found at ${\bf q}$$\sim$$(\pi/2,\pi/2)$
\cite{eder,review}.
Naively, hole pockets at $(\pi/2,\pi/2)$ should appear at finite
hole density. 
In addition, across-the-hole AF bonds are weaker 
at $(\pi/2,\pi/2)$ than at momenta such as ($\pi$,0) 
or ($\pi$,$\pi$)\cite{martins}, although they are still present.
To address this issue let us calculate 
$\rm \langle n_{\bf q} \rangle$=$\rm \langle 
c^\dagger_{\bf q} c_{\bf q} \rangle$, i.e.
the ground-state hole number with 
a given momentum ${\bf q}$ (note that $\rm \langle n_{\bf q} \rangle$ includes 
both coherent and incoherent weight). 
As example, consider
the two-hole problem on the 
4$\times$6 lattice  of Fig.2b. The interesting result in Fig.4a is that 
\vspace{-0.3in}

\begin{figure}
\begin {center}
\mbox{\psfig{figure=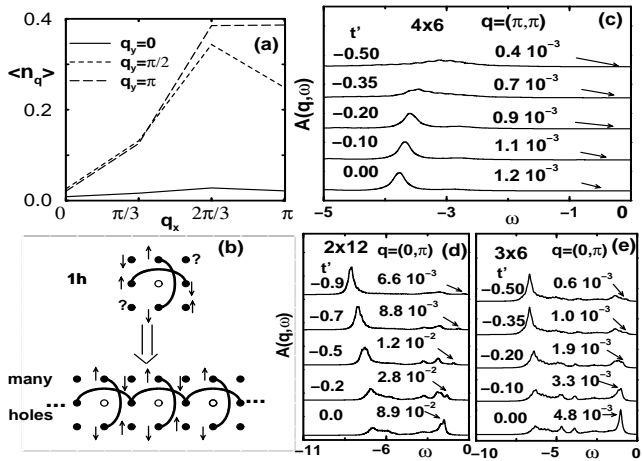,height=2.3in,width=3.2in,rheight=2.5in,rwidth=3.5in,angle=-90}}
\end{center}
\vspace{-0.15in}
\caption{
(a) $\rm \langle n_{\bf q} \rangle$ vs $\rm {\bf q}$, 
for the 2 hole ground-state of
the 4$\times$6 cluster of Fig.2b. (b) 
Qualitative representation of a one-hole state
with strong AF correlations across-the-hole
 as the building-block of n$_h$$\sim$0.5 stripes. In the 1 hole case the
frustration effect is shown with question marks. 
(c) Exact spectral function $\rm A({\bf q},\omega)$  for 
one-hole on the 4$\times$6 cluster with PBC in both
directions, J=0.2, ${\bf q}$=($\pi$,$\pi$) and
t$'$/t$''$=-1.4. Values of t$'$ as well as the (small)
weight in the first pole (and its location) are indicated. 
Note the accumulation of weight at large energies.
(d) Same as (c) but for a 2$\times$12 cluster with PBC 
along legs and ${\bf q}$=(0,$\pi$); (e)
Same as (c) but for a 3$\times$6 cluster (PBC-leg, OBC-rung)
 at ${\bf q}$=(0,$\pi$).} 
\label{fig4}
\end{figure}
\vspace*{-0.1in}
\hspace*{-0.35cm}the ground state 
carries dominant weight at momenta around ($\pi$,$\pi$), and 
the one-hole states with this momentum have robust AF correlations
across-the-hole (Fig.3d), compatible with our
conjecture\cite{dombre}. 
There 
are no indications of small hole-pockets 
in our studies, and the Fermi surface appears open.
In this framework the across-the-hole correlations 
of the, e.g., ($\pi$,$\pi$) or ($\pi$,0) holes can be 
``linked'', as pictorially shown in Fig.4b, improving the hole
mobility since now they share a large region where they do
not need to fight against the spin background to move.
Creating a stripe loop also
avoids the spin frustration intrinsic of the individual hole states
when across-the-hole robust correlations are present (Fig.4b).
In addition, our results help understanding better the
observed stripe density: for n$_h$$\sim$1 the
across-the-hole AF bonds in the stripe direction
cannot form and holes do not improve
their kinetic energy, while for a very hole diluted stripe
the finite-size elementary blocks (Fig.3d) do not touch and cannot 
have a common spin arrangement.
For completeness, in Figs.4c,d the one-hole spectral function
is exactly calculated on 4-, 3- and 2-leg
ladders at small J/t. Note the remarkable
small quasiparticle weight, correlated with a
robust across-the-hole AF correlation (see also \cite{martins}).
The one-hole states contributing to stripes have
exotic properties, including a tendency to spin-charge
separation\cite{martins}.

Summarizing, indications of n$_h$$\sim$0.5
stripes were found
in the extended t-J, t-J$_z$, and 
(at small J/t) in the standard t-J models.
The gain of kinetic energy against the loss of AF energy 
appears enough to stabilize stripes, namely the driving force is
a one hole process and 
the seed for stripes is already present in the insulator.
Contrary to most approaches to stripe formation, here the
{\it small} J/t regime was emphasized.
The scenario reported here is a generalization of the 
1D spin-charge separation involving individual
holons, with the twist that stripes of holons are needed in 
2D to avoid frustration. This result is compatible with Zaanen's
picture of stripes as ``holons in a row''\cite{zaanen}.
Charge and spin could be separated in 2D 
in more subtle ways than anticipated.

The authors thank R. Eder, S. White and J. Zaanen for useful comments and
NSF (DMR-9814350), FAPESP-Brazil, and Fundaci\'on
Antorchas for support.

\vfil


\begin{references}
\bibitem{comm10} Stripes are like many-holes bound states, 
but different from the d-wave pairs of the square lattice.
\bibitem{diagonal}
For the N=18 cluster, ``diagonal'' 3-hole loops
can also be formed.
They carry across-the-stripe $\pi$-shifts and 
substantial ground-state weight.
\end{references}
\end{document}